\documentclass[12pt,a4paper]{article}
\usepackage{epsfig}
\begin{document}

\begin{flushright}
\it
Talk on ACAT2000 Workshop\\
Fermilab, October 16-20, 2000
\end{flushright}

\vspace{1cm}
\begin{center}
\large \bf
Optimization of symbolic evaluation of helicity amplitudes

\end{center}

\vspace{0.5cm}
\begin{center}
P.S.~Cherzor, V.A.~Ilyin and  A.E.~Pukhov
\end{center}
\begin{center}
 \it Skobeltsyn Institute of Nuclear Physics, 
            Moscow State University \\
             Moscow 119899, Russia
\end{center}

\vspace{1cm}

\begin{abstract}
We present a method for symbolic evaluation of Feynman amplitudes.
We construct  special polarization basis for spinor particles which
produces compact expressions for tensor products of basis spinors.
\end{abstract}

\vspace{1cm}


Quantum amplitudes corresponded to Feynman diagrams depend on momenta of
particles and their polarizations. Polarizations of bosons are described by
some Lorentz vectors. Thus, in case, when only boson particles are
involved, the amplitudes may be presented via dot-products and Levi-Civita
tensors.  However fermion polarizations are described by vectors in the
spinor space, so the amplitude is expressed as a product of Dirac
$\gamma$-matrices ended by two spinors (hereafter {\it fermion string}).
Such an expression looks like {\it not complete evaluated} --- on the step
of numerical evaluation one has to perform summations over spinor and
Lorentz indices.  During last years effective algorithms for these
summations were developed, e.g.
\cite{Kleiss,BDK,KS,Tanaka,BalMaina,ALPHA,susyGRACE}, and corresponding
programs were created for automatic computation of Feynman amplitudes, e.g.
\cite{grace,madgraph}\footnote{see also talks by T.~Ohl and C.~Papadopoulos
in these Proceedings \cite{Ohl,HELAC}.}.

Further progress one can get, we believe, if indices summation is performed
before numerical calculations. We refer on our experience with the CompHEP
package \cite{comphep}, where the squared diagram technique is used. For
processes with four particles in final state the number of squared diagrams
becomes very large comparing with the amplitude, and corresponding symbolic
answers are cumbersome. Nevertheless, CompHEP works effectively because it
prepares analytical expression for each squared diagram in terms of
dot-products, and, then, factorizes them. Also, analytical answers for
diagrams can be used to resolve explicitely the cancellations between
diagrams, important for gauge theories.

An idea of the fermion string symbolic evaluation is known long time ago
\cite{FirstTrace}. Although a spinor can not be expressed in terms of
Lorentz vectors, such a possibility exists for pairs of spinors:
\begin{equation}
 w \otimes \bar w' \equiv w^i(p_1,\lambda_1)\bar w'_j(p_2,\lambda_2) \;,
   \label{uu}
\end{equation}
where $p_1$ and $p_2$ are particle momenta, $\lambda_1$ and $\lambda_2$ are
their polarizations. This matrix can be expressed through the Dirac
$\gamma$-matrices convoluted with vectors $p_i$, $\lambda_i$ and some auxiliary
vectors. Thus, the contribution of a fermion string, $\Gamma$, to the Feynman
amplitude can be rewritten through the trace:  
\begin{equation}
  \label{spinor_trace}
    \bar{w}(p_2,\lambda_2) \Gamma \;w'(p_1,\lambda_1)
    = tr(\Gamma \;w\otimes \bar{w}')\;,
\end{equation}
where this string, $\Gamma$, is some product of $\gamma$-matrices.

There are different approaches to the construction of tensor products like
(\ref{uu}), see e.g. \cite{Bondarev, Yehudai,VW}.
Here we propose a new method with the following advantages: 
\begin{enumerate}
\item it produces most compact symbolic expressions for (\ref{uu}); 
\item it gives formulae uniformly applicable in massive and massless cases, for
   Dirac and Majorana fermions;
\item it may be applied to interactions with the fermion number violation; 
\item all phases of fermion states are fixed completely, so one can choose 
   spinor pairs by different ways.
\end{enumerate}

Spinors, corresponded to fermion external legs of the diagram, can be one
of the following variants, $ w \in  \{ u,v,u^c,v^c \}$ and $\bar{w}' \in
\{ \bar{u}, \bar{v}, \overline{u^c}, \overline{v^c} \} $. Here $(^c)$
denotes the operation of C-conjugation, the corresponding fermion strings
appear, for example, in the case of interaction with fermion number violation
\cite{Denner}. Spinors $u$ and $v$ are defined as
\begin{eqnarray}
\label{u_def}
<0|\psi(x)|\Omega_{+}(p,\lambda)>&=& e^{-i p x}u(p,\lambda)\;, \\
\label{v_def}
<0|\bar{\psi}(x)|\Omega_{-}(p,\lambda)>&=& e^{-i p x} \bar{v}(p,\lambda)\;,
\end{eqnarray}
where $|0>$  is vacuum state, $\Omega_{+}$ ($\Omega_{-}$) is one-particle
fermion (anti-fermion) state. The fermion states depend on
the choice of the polarization axises and their complex phases are not fixed.

The characteristic property of spinors $u$ and $v$ is the {\it completeness}:
\begin{eqnarray}
\label{completeness1}
\Sigma_\lambda \; u(p,\lambda)\otimes \bar{u}(p,\lambda)&=&\hat{p}+m\\
\label{completeness2}
\Sigma_\lambda \; v(p,\lambda) \otimes \bar{v}(p,\lambda)&=&\hat{p}-m
\end{eqnarray}
where $m=\sqrt{p^2}$. One can say that any spinors, $u(p,\lambda)$ and
$v(p,\lambda)$, satisfied these relations correspond to some polarization
states $\Omega_{+}$ and $\Omega_{-}$.

By definition, for Majorana fermion one has
$\psi(x)=\psi(x)^c$ and $\Omega_{+}(p,\lambda)=\Omega_{-}(p,\lambda)$.
If substitute these identities into 
(\ref{u_def},\ref{v_def}) one gets
\begin{equation}
\label{u_to_v}
u(p,\lambda) = C \bar{v}(p,\lambda)^T  = v(p,\lambda)^c \;.
\end{equation}
Thus, we conclude that spinor basis should satisfy relation (\ref{u_to_v})
in order to construct an algorithm for evaluation of fermion strings with
Majorana fermions. Hereafter we assume that spinors $u$ and $v$ satisfy this
relation.

Following to \cite{Kleiss,BDK} we construct $u$ and $v$ spinors in arbitrary
point of the momentum space as a projection of spinors defined at some
auxiliary point $p_0$:
\begin{eqnarray}
u(p,\lambda)&=& N (m+\hat{p}) u(p_0,\lambda)\;,\\
v(p,\lambda)&=& N (m-\hat{p}) v(p_0,\lambda)\;,
\end{eqnarray}
where $N=1/\sqrt{2[(p_0p)+m_0 m]}$. It is easy to check that this procedure
agrees with the completeness  (\ref{completeness1}, \ref{completeness2}) at
any point $p$ if it is  satisfied at the point $p_0$. Also, it is
enough  to satisfy (\ref{u_to_v}) at $p_0$ to get this relation at any 
point.

Let $p_0$ is massless vector, in this case expressions appeared are
simplified. In particular, equations (\ref{completeness1}) and
(\ref{completeness2}) become to be identical,  and, thus, (massless) $u$
and $v$ spinors are basis vectors in the same spinor subspace. It is well
known, that left-handed and right-handed spinors,
$\xi_{-}=-\gamma_5\xi_{-}$ and $ \xi_{+}=\gamma_5\xi_{+}$, may be chosen as
an elements of this basis, and completeness reads as $ \xi_{-} \otimes
\bar{\xi}_{-} + \xi_{+} \otimes \bar{\xi}_{+}=\hat{p}_0 $. Note that chiral
spinors, $\xi_{\pm}$, are defined by these relations up to phases. Let us
fix these phases.

At first, let us connect chiral spinors with massless $v$ spinor of different
polarizations taken at the point $p_0$:
$$\xi_{+}=v(p_0,+1)\;, \qquad \xi_{-}=v(p_0,-1)\;.
$$
Then, from (\ref{u_to_v}) one gets
$$
  u(p_0,+1)=\xi_{+}^c\;, \qquad
  u(p_0,-1)=\xi_{-}^c\;.
$$
Note, then, that C-conjugation transforms the subspace of
$u$-spinors to the subspace of $v$-spinors. For massless $p_0$ these
subspaces are identical and, thus, invariant under the C-conjugation. From
other side C-conjugation transforms left-handed spinor to right-handed
spinor. So, sum of the phases of $\xi_{\pm}$ can be fixed by
the relation
\begin{equation}
  \label{lCr}
    \xi_{-}=\xi_{+}^c\;.
\end{equation}

As a result of our construction the
following expressions for $u$ and $v$ spinors can be derived:
\begin{equation}
\label{res1_1}
u(p,\pm 1)=v(p,\pm 1)^c = N (m+\hat{p})\xi_{\mp}\,,
\end{equation}
\begin{equation}
\label{res1_2}
v(p,\pm 1)=u(p,\pm 1)^c = N (m-\hat{p})\xi_{\pm}\,,
\end{equation}
where $N=1/\sqrt{2(p_0p)}$.

By means of Dirac conjugation one immediately gets
\begin{equation}
\label{res2_1}
\bar{u}(p,\pm 1)=\overline{v(p,\pm 1)^c} =
   N \bar{\xi}_{\mp}  (m+\hat{p})\,, 
\end{equation}
\begin{equation}
\label{res2_2}
\bar{v}(p,\pm 1)=\overline{u(p,\pm1)^c} =
   N \bar{\xi}_{\pm}(m-\hat{p})\,.
\end{equation}

Expressions (\ref{res1_1} - \ref{res2_2}) reduce the problem  of evaluation
of matrices $w \otimes \bar{w}'$ to the evaluation of 
$\xi_{+}\otimes\bar{\xi}_{+}$, $\xi_{-}\otimes\bar{\xi}_{-}$,
$\xi_{+}\otimes\bar{\xi}_{-}$ and $\xi_{-}\otimes\bar{\xi}_{+}$.  The
first two products can be derived directly from
the definition of chiral spinors:
\begin{eqnarray}
\label{norm_xi_plus}
    \xi_{+}\otimes\bar{\xi}_{+}&=& \frac{1+\gamma_5}{2}\hat{p}_0 \;,\\
\label{norm_xi_minus}
    \xi_{-}\otimes\bar{\xi}_{-}&=& \frac{1-\gamma_5}{2}\hat{p}_0 \;.
\end{eqnarray}

However, expressions for last two products, $\xi_{+}\otimes\bar{\xi}_{-}$
and  $\xi_{-}\otimes\bar{\xi}_{+}$, can not be derived yet, because they
are sensitive to the multiplication of $\xi_{\pm}$ by opposite phase
factors. Note that rotation of 3-dimension coordinate system around the
space component of $p_0$, will produce the multiplication of $\xi_{\pm}$ 
by opposite phase factors. Thus, one needs to introduce addition auxiliary
vectors to fix the reference frame, and (as a result) to fix phases, and
(finally) to get a representation for $\xi_{+}\otimes\bar{\xi}_{-}$ and
$\xi_{-}\otimes\bar{\xi}_{+}$. Let $\eta_1$ and  $\eta_2$ are space-like
vectors orthogonal to $p_0$:
\begin{equation}
\label{eta_norm}
   \eta_1.\eta_2= 0\,,\quad
   \eta_i.p_0=0\,,\quad
   \eta_i.\eta_i=-1\,.
\end{equation}
Three vectors ($p_0$, $\eta_1$, $\eta_2$) may be completed to  the
basis in Minkowsky space \footnote{say, by vector $q_0$, such that 
$q_0.p_0=1$ and  $q_0.\eta_i=0$.}. Now we need that this basis is
right-handed oriented. It can be done by fixing the sign of the convolution
of Levi-Civita tensor with basis vectors. We fix the basis orientation by
the relation
\begin{equation}
\label{orientation_spin}
\hat{p}_0=i\gamma_5\hat{\eta}_1\hat{\eta}_2\hat{p}_0\;.
\end{equation}
If one constructs complex vectors 
$ \eta \equiv (\eta_1 + i \eta_2)/2$ and
$\eta^*\equiv (\eta_1 - i \eta_2)/2$,
then (\ref{orientation_spin}) can be rewritten as
\begin{equation}
  \label{ceta_orientation}
  \hat{\eta}^*\hat{\eta}\hat{p}_0=-\frac{1-\gamma_5}{2} \hat{p}_0 \;.
\end{equation}
Using this equation one can check that
\begin{equation}
\label{plusTominus}
\xi_{-}=\hat{\eta}^*\xi_{+}\;.
\end{equation}
Note that (\ref{norm_xi_minus}) defines $\xi_{-}$ uniquely up to a phase.
Thus (\ref{plusTominus}) fixes the difference of phases between
$\xi_{-}$ and $\xi_{+}$ (whereas (\ref{lCr}) fixes their sum). 
Now one can derive from (\ref{plusTominus}) the second pair of products:
$$
  \xi_{-}\otimes\bar{\xi}_{+}= \frac{1-\gamma_5}{2}\hat{\eta}^*\hat{p}_0 \;, 
\quad
  \xi_{+}\otimes\bar{\xi}_{-}=
-\frac{1+\gamma_5}{2}\hat{\eta}\hat{p}_0 \;.
$$
Note that these relations are some variant of the
Bouchiat-Michel equations \cite{BM,haber}.

Now we can write down the complete set of expressions for tensor products
$w\otimes\bar{w}'$:
\begin{eqnarray}
\label{uu_answer}
u(p,\lambda)\otimes \bar{u}(p',\lambda') &=& N (m+\hat{p})
\left( \begin{array}{cc} 
           1-\gamma_5              & (1-\gamma_5)\hat{\eta}^* \\
          -(1+ \gamma_5)\hat{\eta} & 1+\gamma_5
       \end{array}
\right)   \hat{p}_0 (m'+\hat{p}') \,;   \\
\label{uv_answer}
u(p,\lambda)\otimes \bar{v}(p',\lambda') &=& N (m+\hat{p})
\left( \begin{array}{cc}
         (1- \gamma_5)\hat{\eta}^* & 1-\gamma_5 \\
          1+\gamma_5               & -(1+\gamma_5)\hat{\eta}
       \end{array}
\right)  \hat{p}_0 (m'-\hat{p}')\,;    \\
\label{vu_answer}
v(p,\lambda)\otimes \bar{u}(p',\lambda') &=& N (m-\hat{p})
\left( \begin{array}{cc}
          -(1+ \gamma_5)\hat{\eta} & 1+\gamma_5 \\ 
           1-\gamma_5              & (1-\gamma_5)\hat{\eta}^*
       \end{array}
\right) \hat{p}_0(m'+\hat{p}') \,;  \\
\label{vv_answer}
v(p,\lambda)\otimes \bar{v}(p',\lambda') &=& N (m-\hat{p})
\left( \begin{array}{cc} 
          1+\gamma_5               & -(1+\gamma_5)\hat{\eta} \\
         (1- \gamma_5)\hat{\eta}^* & 1-\gamma_5
       \end{array}
\right) \hat{p}_0(m'-\hat{p}') \,. 
\end{eqnarray}
Here $N=1/[4\sqrt{(p p_0)(p'p_0)}]$, $m=\sqrt{p^2}$, $m'=\sqrt{p'^2}$, 
and dependence on the polarizations follows the rule:
{\small $\left( \begin{array}{cc}
           ++ & +- \\
           -+ & --
        \end{array}
\right)
$. }
The products $w\otimes\bar{w}'$  with C-conjugation may be derived now by
means of relation (\ref{u_to_v}).

Note, that for $m \neq 0$ the spinors $u$ and $v$ have spin directed along
the axis $n=\frac{1}{m}p - \frac{m}{(pp_0)}p_0$. In massless case spinors
$u(p,+1)$ and $v(p,-1)$ are right-handed, whereas $u(p,-1)$ and $v(p,+1)$
are left-handed.

At this step one can conclude that in the proposed method the same set of
auxiliary vectors ($p_0$, $\eta$, $\eta^*$) is introduced for arbitrary
number of fermion strings in the amplitudes. Moreover, only one set can be
used for all diagrams to fix relative phases of spinors for interference
diagrams. One can compare with other methods where number of sets of
auxiliary vectors depend on the number of fermion strings (see, e.g.,
corresponding discussion in \cite{Bondarev}). 

One can find further optimizations in the proposed technique. Indeed,
vectors $\eta$ and $\eta^*$ in (\ref{uu_answer}-\ref{vv_answer}) are
accompanied always by the $p_0$ vector. Let us define antisymmetric tensor
$G_{\mu\nu}\equiv -i\epsilon_{\mu\nu\alpha\beta}\eta^{\alpha}p_{0}^{\beta}$. Then
one can check that 
$
\hat{\eta}\hat{p}_0 = \frac{1}{2}G_{\mu\nu}\gamma^{\mu}\gamma^{\nu}
$ 
and
$\hat{\eta}^*\hat{p}_0 = \frac{1}{2}G^*_{\mu\nu}\gamma^{\mu}\gamma^{\nu}
$.
By substituting these relations in (\ref{spinor_trace}) one can hide the
dependence on $\eta$, $\eta^*$ inside the $G$ tensor. 

The $G$ tensor has some properties allowed one to perform the indices
convolutions:
$$
G_{\mu\nu}{p_0}^{\mu}=0\;,\quad
G_{\mu\nu}{G_{\mu'}}^{\nu}=0\;,\quad
{G_{\mu}}^{\nu}G^*_{\nu\mu'}=2{p_0}_{\mu}{p_0}_{\mu'}\,,
$$
$$
g^{\mu\mu'}G_{\mu\nu}\epsilon_{\mu'\alpha\beta\gamma}=
4i(g_{\nu\alpha}
G_{\beta\gamma}-g_{\nu\beta}
G_{\alpha\gamma}+g_{\nu\gamma}
G_{\alpha\beta})\;.
$$
As a result, one can get symbolic answer for the amplitude expressed via
scalar products of particle momenta and auxiliary vector $p_0$, their
convolutions with the Levi-Civita tensor, and particle momenta convolved
with the $G$ tensor. The use of $G$ tensor optimizes the final expressions ---
if expand $G$ tensor via scalar products two terms appear.

Then, one can expect that amplitude contains both $G$ and $G^*$ tensors.
However, the product of these tensors,  $G(q_1,p_1)G^*(q_2,p_2)$, can be
expressed via the dot-products and Levi-Civita tensors. So, one can
connects fermion legs of the diagram in such a way that amplitudes will be
presented via either $G$ or $G^*$ tensors, but never both together. 

This work was done in the framework of the CPP Collaboration \cite{CPP},
and supported by grants INTAS-CERN 99-0377 and  INTAS 96-842. A.E.P. was
supported also by the Programme "University of Russia" (grant 990588).

\nocite{*}
\bibliographystyle{aipproc}

\end{document}